\documentclass[pra, twocolumn, final, amsmath, amssymb, superscriptaddress, a4paper, aps, 10pt, hidelinks, numbers, sort&compress]{revtex4-1}
\usepackage{amsmath}%
\usepackage{amsfonts}%

\usepackage{mathptmx, textcomp}
\usepackage[latin1]{inputenc}
\usepackage[usenames]{color}
\usepackage{graphicx}           
\usepackage{dcolumn}            
\usepackage{bm}                 
\usepackage{amssymb, amsmath}   
\usepackage[usenames]{color}
\usepackage[colorlinks,urlcolor=black,citecolor=blue,linkcolor=black]{hyperref}
\begin{document}
\newcommand{\bra}[1]{\mbox{\ensuremath{\langle #1 \vert}}}
\newcommand{\ket}[1]{\mbox{\ensuremath{\vert #1 \rangle}}}
\newcommand{\mb}[1]{\mathbf{#1}}
\newcommand{\phipp}{\big|\phi_{\mb{p}}^{(+)}\big>}
\newcommand{\phipav}{\big|\phi_{\mb{p}}^{\p{av}}\big>}
\newcommand{\pp}[1]{\big|\psi_{p}(#1)\big>}
\newcommand{\drdy}[1]{\sqrt{-R'(#1)}}
\newcommand{\Rb}{$^{87}$Rb}
\newcommand{\K}{$^{40}$K }
\newcommand{\Li}{$^{6}$Li }
\newcommand{\LiK}{$^6$Li-$^{40}$K}
\newcommand{\na}{${^{23}}$Na}
\newcommand{\muK}{\:\mu\textrm{K}}
\newcommand{\p}[1]{\textrm{#1}}
\newcommand\T{\rule{0pt}{2.6ex}}
\newcommand\B{\rule[-1.2ex]{0pt}{0pt}}
\newcommand{\reffig}[1]{\mbox{Fig.~\ref{#1}}}
\newcommand{\refeq}[1]{\mbox{Eq.~(\ref{#1})}}
\newcommand{\1}{\ensuremath{\left|1 \right\rangle}}
\newcommand{\2}{\ensuremath{\left|2 \right\rangle}}
\newcommand{\3}{\ensuremath{\left|3 \right\rangle}}

\hyphenation{Fesh-bach}
\newcommand{\previous}[1]{}
\newcommand{\fs}[1]{\textcolor{red}{[#1]}}
\newcommand{\gboxx}{\textcolor{ForestGreen}{$\Box$}}
\newcommand{\bcirc}{\textcolor{blue}{$\bigcirc$}}
\newcommand{\cit}{\textcolor{red}{[cite]}}
\newcommand{\bk}{{\bf k}}
\newcommand{\beq}{\begin{equation}}
\newcommand{\eeq}{\end{equation}}
\newcommand{\bei}{\begin{itemize}}
\newcommand{\eei}{\end{itemize}}
\newcommand{\ben}{\begin{enumerate}}
\newcommand{\een}{\end{enumerate}}
\newcommand{\nn}{\nonumber}
\newcommand{\bp}{{\bf p} }
\newcommand{\bq}{{\bf q} }
\newcommand{\br}{{\bf r} }
\newcommand{\down}{\downarrow}
\newcommand{\up}{\uparrow}
\newcommand{\cT}{{\mathcal T}}

\setcounter{figure}{0}
\setcounter{equation}{0}
\setcounter{section}{0}
\setcounter{table}{0}

\title{Exploring emergent heterogeneous phases in strongly repulsive Fermi gases}
\author{F. Scazza}
\affiliation{Istituto Nazionale di Ottica del Consiglio Nazionale delle Ricerche (CNR-INO), 50019 Sesto Fiorentino, Italy}
\affiliation{\mbox{European Laboratory for Non-Linear Spectroscopy (LENS), Universit\`{a} di Firenze, 50019 Sesto Fiorentino, Italy}}
\author{G. Valtolina}
\affiliation{JILA, University of Colorado, Boulder, CO 80309, USA}
\author{A. Amico}
\affiliation{Istituto Nazionale di Ottica del Consiglio Nazionale delle Ricerche (CNR-INO), 50019 Sesto Fiorentino, Italy}
\affiliation{\mbox{European Laboratory for Non-Linear Spectroscopy (LENS), Universit\`{a} di Firenze, 50019 Sesto Fiorentino, Italy}}
\affiliation{\mbox{Dipartimento di Fisica e Astronomia, Universit\`{a} di Firenze, 50019 Sesto Fiorentino, Italy}}
\author{P. E. S.~Tavares}
\affiliation{Departamento de F\'isica, Universidade Federal de Minas Gerais, CEP 31270-901 Belo Horizonte, MG, Brasil}
\author{M. Inguscio}
\affiliation{Istituto Nazionale di Ottica del Consiglio Nazionale delle Ricerche (CNR-INO), 50019 Sesto Fiorentino, Italy}
\affiliation{\mbox{European Laboratory for Non-Linear Spectroscopy (LENS), Universit\`{a} di Firenze, 50019 Sesto Fiorentino, Italy}}
\affiliation{Department of Engineering, Campus Bio-Medico University of Rome, 00128 Rome, Italy}
\author{W. Ketterle}
\affiliation{Department of Physics, MIT-Harvard Center for Ultracold Atoms, and Research Laboratory of Electronics, MIT, Cambridge, Massachusetts 02139, USA}
\author{G. Roati}
\affiliation{Istituto Nazionale di Ottica del Consiglio Nazionale delle Ricerche (CNR-INO), 50019 Sesto Fiorentino, Italy}
\affiliation{\mbox{European Laboratory for Non-Linear Spectroscopy (LENS), Universit\`{a} di Firenze, 50019 Sesto Fiorentino, Italy}}
\author{M. Zaccanti}
\email[Corresponding author. E-mail: ]{matteo.zaccanti@ino.cnr.it}
\affiliation{Istituto Nazionale di Ottica del Consiglio Nazionale delle Ricerche (CNR-INO), 50019 Sesto Fiorentino, Italy}
\affiliation{\mbox{European Laboratory for Non-Linear Spectroscopy (LENS), Universit\`{a} di Firenze, 50019 Sesto Fiorentino, Italy}}

\begin{abstract}
Recent experiments have revitalized the interest in  a  Fermi gas of ultracold atoms with strong repulsive interactions. 
In spite of its seeming simplicity, this system exhibits a complex behavior, resulting from the competing action of two distinct instabilities: ferromagnetism, which promotes spin anticorrelations and domain formation; and pairing, that renders the repulsive fermionic atoms unstable towards forming weakly bound bosonic molecules. 
The breakdown of the homogeneous repulsive Fermi liquid arising from such concurrent mechanisms has been recently observed in real time through pump-probe spectroscopic techniques [A.~Amico \textit{et al.}, Phys. Rev. Lett. \textbf{121}, 253602 (2018)]. These studies also lead to the discovery of an emergent metastable many-body state, an unpredicted quantum emulsion of anticorrelated fermions and pairs. Here, we investigate in detail the properties of such an exotic regime by studying the evolution of kinetic and release energies, the spectral response and coherence of the unpaired fermionic population, and its spin-density noise correlations. All our observations consistently point to a low-temperature heterogeneous phase, where paired and unpaired fermions macroscopically coexist while featuring micro-scale phase separation. Our findings open new appealing avenues for the exploration of quantum emulsions and also possibly of inhomogeneous superfluid regimes, where pair condensation may coexist with magnetic order.  
\end{abstract}

\maketitle

\section{Introduction}

In strongly correlated electron systems, such as transition-metal oxides and heavy fermion compounds, the simultaneous presence of 
multiple interaction mechanisms 
and the concurrence of distinct competing instabilities can lead to the spontaneous emergence of spatially inhomogeneous states. These are characterized by nanometer-scale structures hosting different phases and order parameters \cite{Dagotto2005, Dagotto2013}. The complexity of such materials lies at the heart of remarkable effects such as colossal magnetoresistance, and it may play a crucial role even for unconventional  superconductivity \cite{Dagotto2003}. Furthermore, it might have important consequences for applications of such materials beyond semiconducting electronics and spintronics, with the availability of different degrees of freedom simultaneously active in the many-body system, enabling giant responses to small perturbations \cite{Dagotto2005}. 

At a first glance, the great complexity of electron matter starkly contrasts with the simplicity offered by ultracold Fermi gases. Owing to their exceptional cleanliness and tunability, such systems are indeed considered as ideal testbeds for studying strong correlation phenomena within a variety of minimal many-body Hamiltonians. In particular, two-component Fermi mixtures with attractive interaction have been widely employed to experimentally explore the BCS-BEC crossover, providing fundamental new insights into the universal unitary regime \cite{Varenna2007, Zwerger2011}, also for imbalanced spin populations \cite{Chevy2010, Radzihovsky2010, Massignan2014}.
On the other hand, repulsive Fermi gases close to a Feshbach resonance have been originally regarded as an ideal platform for investigating the textbook Stoner model for itinerant ferromagnetism \cite{Stoner1933, Salasnich2000, Sogo2002, Duine2005, Jo2009}, as this system  embodies the two ingredients of Stoner's Hamiltonian -- Fermi pressure and short-range repulsion -- free from intricate band structures, additional kinds of interactions and disorder inherent to any condensed matter system. 

However, experimental attempts in this direction revealed a much richer and more subtle behavior, very different from the original Stoner scenario. Although convincing signatures for a ferromagnetic instability within the repulsive Fermi liquid have been obtained through  studies of spin dynamics \cite{Valtolina2017} and time-resolved quasi-particle spectroscopy \cite{Scazza2017, Amico2018}, already early experiments \cite{Jo2009,Sanner2012} found the system dynamics to be fundamentally affected by another type of instability, antithetical to ferromagnetism and associated with the tendency of repulsive fermions to combine into weakly bound pairs. This latter mechanism is intimately linked to the short-ranged nature of the interatomic interaction: The strong repulsion necessary for ferromagnetism to develop can only be attained if a weakly bound molecular state exists below the two-atom scattering threshold. As such, the repulsive Fermi gas represents an excited metastable branch of the many-body system, inherently affected by decay processes towards lower-lying molecular states,  which become faster for larger repulsive interactions \cite{Shenoy2011, Pekker2011, Sanner2012}.

Pump-probe spectroscopic studies \cite{Amico2018} recently measured the rates at which pairing and short-range ferromagnetic correlations develop after creating a strongly repulsive Fermi liquid, finding the two instabilities to rapidly grow over comparable timescales. Moreover, the same survey revealed that both mechanisms persist at long times, leading to a semi-stationary regime consistent with a spatially heterogeneous phase. A natural framework to understand this state is a quantum emulsion, where pairs and unpaired fermions macroscopically coexist while featuring phase segregation at the micro-scale of few interparticle spacings \cite{Amico2018}. 
While this complex behavior undermines the prospects for the realization of the basic Stoner model with ultracold atoms, it connects the fate of repulsive Fermi gases to other intriguing instances of highly-correlated fermionic matter, and it thus makes their exploration highly relevant \textit{per se}. In particular, the coexistence of  paired and magnetic states within an ultracold quantum emulsion \cite{Roscilde2007} could represent an unforeseen opportunity, as it possibly embodies the simplest form of complexity within a strongly-correlated fermionic system. The existence of such heterogeneous state holds promise for novel phenomena, encompassing anomalous transport and localization \cite{Roscilde2007, Ospelkaus2006, Sanchez-Palencia2010}, non-trivial and yet poorly explored fermion-pair interactions \cite{Jag2014, Kinnunen2018, Camacho-Guardian2018}, and exotic superfluid phases \cite{Bulgac2006}.

In this work, we provide a detailed characterization of the various properties of this long lived quantum emulsion regime of a two-component Fermi mixture of ultracold $^6$Li atoms. In particular, by complementing high-resolution spectroscopy measurements \cite{Amico2018} with trap release experiments and controlled interaction or temperature changes, we gain new insight into the development of ferromagnetic correlations and their persistence in the presence of pairing, which acts as a competing  channel. By inspecting the quasi-particle coherence properties of the unpaired atomic components within the emerging fermion-pair mixture, we observe the dynamical breakdown of the repulsive Fermi liquid associated with a real-time suppression of the quasi-particles dressing, pointing to a vanishing short-range overlap between the two spin components. 
Finally, new spectroscopic and noise-correlation measurements provide further indications about the spatially heterogeneous character of such exotic many-body state.

The paper is organized as it follows: In Section 2 we recall the experimental procedure to selectively prepare the initial repulsive Fermi liquid state through rapid radio-frequency transfers. 
Section 3 presents evidence for the  breakdown of the repulsive Fermi liquid and its relation to Stoner ferromagnetism, based on the observed dynamics of kinetic and interaction energies, and quasi-particle coherence measurements. Section 4 reports on the metastable, low-temperature nature of the correlated state developing after long hold times. Section 5 discusses spin density fluctuation and correlation measurements, and additional spectroscopic probing of the system, providing compelling arguments in favour of the spatially heterogeneous character of the emulsion state.

\vspace{-0.2cm}
\section{Selective Preparation of a strongly Repulsive Fermi liquid state}

\begin{figure}[t!]
\centering
\includegraphics[width= 7.6 cm]{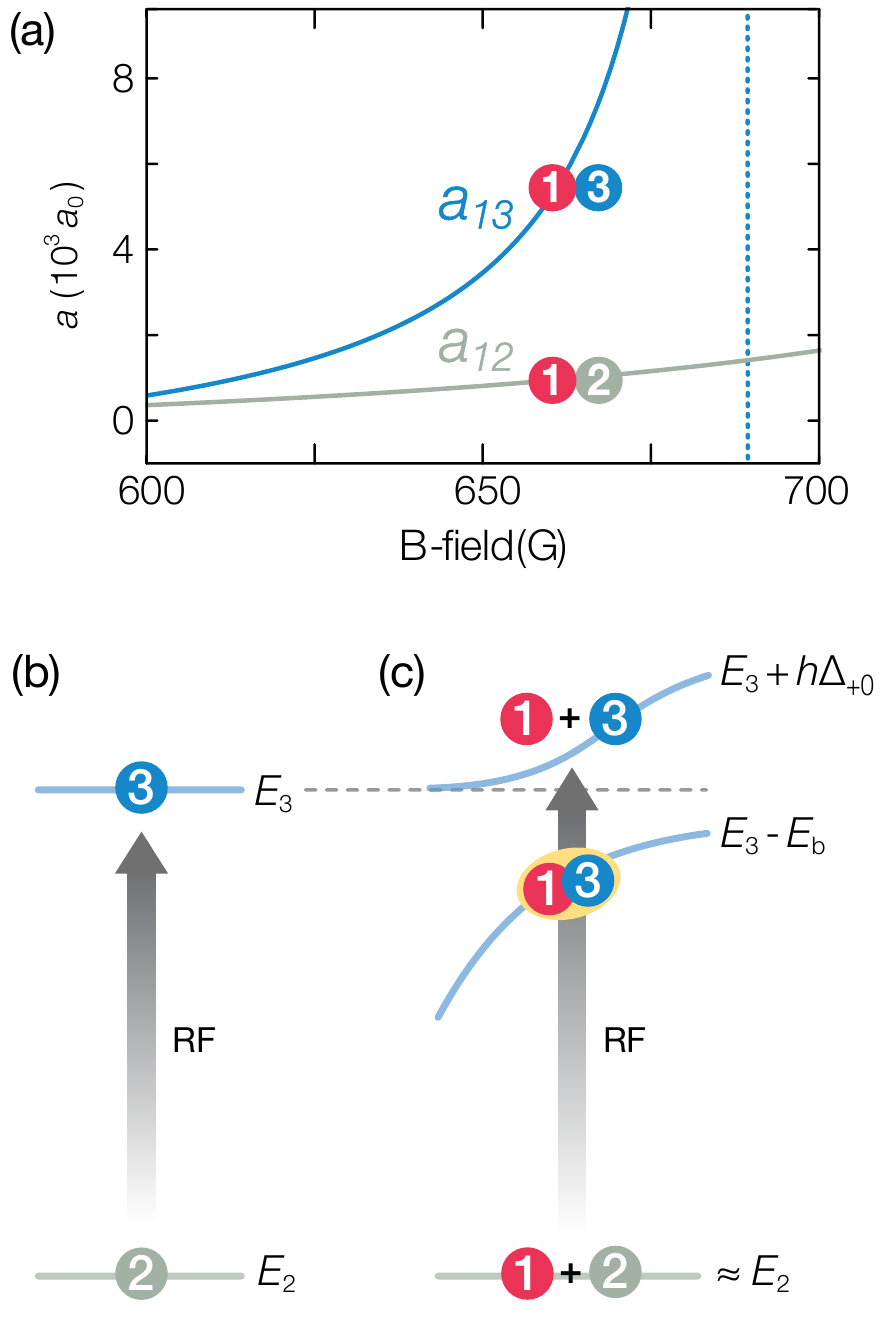}
\caption{
\label{figquenchbranch} 
Protocol to prepare a strongly repulsive Fermi liquid. (a) Scattering lengths $a_{12}$ and $a_{13}$ between the 1-2 (grey) and 1-3 (blue) atomic states, respectively, as a function of the bias magnetic field strength. Two well-separated Feshbach resonances at 832\,G and 690\,G allow for tuning $a_{13}$ to large values, while only weakly affecting $a_{12}$. (b) A spin-polarized Fermi gas of state-2 atoms is transferred in the adjacent state 3 by a radio-frequency photon with frequency $\nu_0 \simeq 83$\,MHz, matching the bare atomic energy splitting $h \nu_0 = E_3 - E_2$. (c) When mixed together with state-1 atoms, the spectral response is strongly modified by short-range 1-3 interactions. At magnetic fields between 640 and 680\,G, the associated $2 \rightarrow 3$ spectrum is characterized both by a repulsive atomic peak located at positive detuning $\Delta_{+0}$ from $\nu_0$, and a separate incoherent molecular association spectrum located at frequencies lower than $\nu_0$. Selective transfer along the upper branch of the Feshbach resonance, i.e.~conversion of a weakly repulsive 1-2 mixture into a strongly repulsive 1-3 one, is obtained by applying a $\sim100\,\mu$s-long RF pulse at positive detuning $\Delta_{+0}$. 
}
\end{figure}

We produce a weakly interacting  mixture of $N \simeq 2 \times 10^5$ of  $^6$Li atoms, equally populating the two lowest hyperfine states, denoted as 1 and 2, respectively \cite{Burchianti2014}. The atomic cloud is held in a cylindrically shaped optical dipole trap at a temperature as low as $T \simeq 0.1\,T_F$, where $T_F = E_F/k_B$ is the Fermi temperature and $k_B$ is the Boltzmann constant. $E_F=\hbar (6 N \omega_x \omega_y \omega_z)^{1/3}$ denotes the peak Fermi energy, with $\hbar$ being the reduced Planck constant $h/(2\pi)$. 
In order to reduce the effects of density inhomogeneity, we typically record the spectral response only within a central region of the trapped cloud \cite{Amico2018}. The initial average density of state-2 atoms within this region sets the relevant Fermi energy $\epsilon_F$ and wavevector $\kappa_F$ of the gas. This in turn sets a minimum Fermi time $\tau_F=h/\epsilon_F \simeq 20\,\mu$s.
Selective and rapid transfer to strong interactions is achieved by following a radio-frequency (RF) spin-injection protocol analogous to that presented in Ref. \cite{Amico2018}, driving the transition from state 2 to state 3, i.e. the third-to-lowest hyperfine state. The broad Feshbach resonance between states 1 and 3 located at a bias magnetic field of 690\,G enables to controllably tune the 1-3 $s$-wave scattering length $a$, and hence the contact interaction strength parametrized by $\kappa_Fa$. 
Within the magnetic field range explored in this study, spanning between 640 and 680\,G, the 1-3 mixture is strongly repulsive, whereas any other pairwise combination involving the three lowest Zeeman states features only weak interactions [see Fig.~\ref{figquenchbranch}a]. 
By setting the RF field at a detuning corresponding to the repulsive quasi-particle resonance [see Fig.~\ref{figquenchbranch}b-c], previously determined via precision RF spectroscopy and hereafter denoted as $\Delta_+(0)$, we selectively convert weakly repulsive 1-2 mixtures into strongly repulsive 1-3 ones by means of RF $\pi$-pulses with a duration of about $150\,\mu$s. Depending on the final 1-3  interaction, we obtain $2 \rightarrow 3$ transfer efficiencies that typically range between 98$\%$ and 75$\%$ for the weakest and strongest repulsion explored here, respectively. 
If not otherwise specified, the remaining state-2 atom population is selectively expelled from the trap immediately after the end of the RF pulse by means of a a 3\,$\mu$s-long resonant optical blast, that negligibly affects the 1-3 mixture.

At the end of this procedure, the weakly interacting 1-2 mixture has been converted into a 1-3 strongly repulsive Fermi liquid, whose subsequent many-body dynamics can be monitored through different protocols, both at short and long evolution times. In Ref. \cite{Amico2018}, we have reported on pump-probe spectroscopic measurements that enabled us to quantify both the initial growth rate of ferromagnetic and pairing correlations, and their competing action towards the reach of a semi-stationary regime. In essence, such a scheme relies on the study of both the center frequency  $\Delta_+(t) \geq 0$, amplitude $A(t)$ and width $w_{+}(t)$ of the coherent peak associated with the unpaired repulsive fermion population identified in the $3 \rightarrow 2$ probe response. At each time during the post-quench evolution, $\Delta_+(t)$ quantifies the interaction energy characterizing state-3 atoms within the $1-3$ mixture, proportional in turn to the short-range 1-3 pair correlations \cite{Amico2018}. The atomic peak area, proportional to $A(t) \cdot w_{+}(t)$, owing to the absence of an overall density drop within the observation region, provides instead a real-time measure of the state-3 atomic population that has not recombined into molecular pairs. Finally, the atomic peak width $w_+(t)$ provides information about the \textit{coherence} of the fermionic states, and about both thermal and collisional effects \cite{Amico2018}.   
In the present study we combine such pump-probe spectroscopic technique with controlled interaction changes, implemented through magnetic field sweeps, as well as with other probing methods. These include measurements of kinetic and release energy, characterization of the atom-pair mixture stability, and studies of short-range spin density noise and correlations. 
\begin{figure}[!b]
\centering
\includegraphics[width= 7.5 cm]{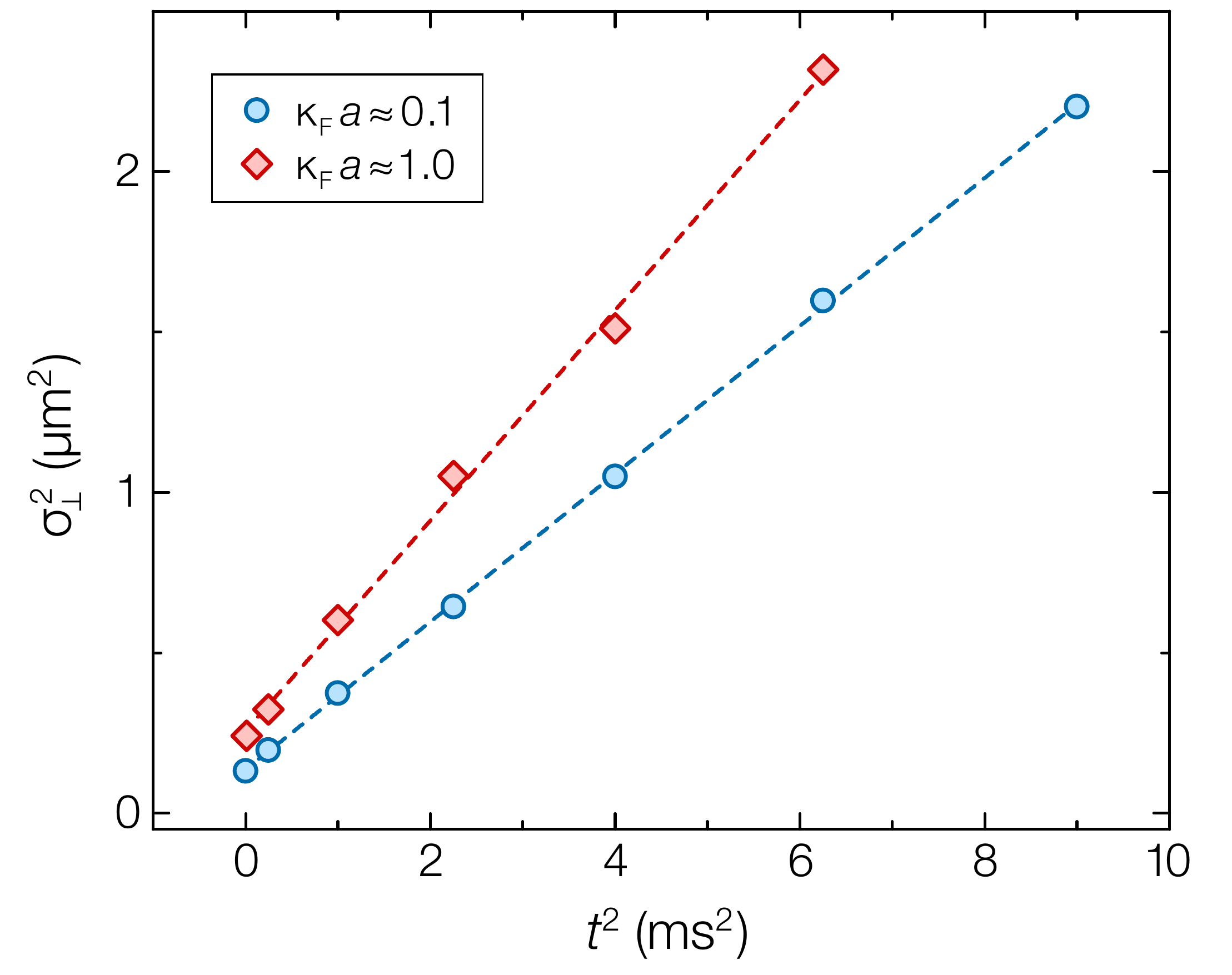}
\caption{\label{FigEkin_TOF}
Examples of radial width measurements in time-of-flight expansion for a weakly interacting 1-3 mixture at $k_Fa=0.1$ (blue circles) and a strongly repulsive one at $k_Fa=1.0$ (red diamonds). Data are recorded after 2\,ms of evolution from the RF transfer. The data are plotted using a quadratic scale both for the time of flight and the radial width, rendering the slope proportional to the kinetic energy. All data points are measured at $T/T_F = 0.12(2)$.
}
\vspace*{0pt}
\end{figure}

\vspace{-0.3cm}
\section{Characterization of the atomic domains within an atom-pair emulsion}
\vspace{-0.2cm}

Previous studies on  repulsive Fermi gases following an interaction quench revealed a behavior  of the mean kinetic energy of the particles which was not monotonic as a function of the positive scattering length, increasing sharply or critical values of repulsion \cite{Jo2009}.
 This was interpreted as an indirect signature for a Stoner-like instability of the paramagnetic Fermi liquid towards formation of spin-polarized domains.
 When this happens, the short-range interaction energy is drastically reduced, at the cost of an increased kinetic energy associated with the larger Fermi pressure of a more tightly confined sample. However, later experiments by the same group \cite{Sanner2012} found that only part of the total energy was indeed converted into kinetic energy, the latter remaining a minor fraction of the release energy.  Combined with the observation of a substantial and rapid conversion of the atomic cloud into paired states, this suggested the absence of a ferromagnetic phase. Here we show that the situation is actually richer: A quantum emulsion of atoms and pairs combines properties of a ferromagnetic and a paired state.
%
\subsection{Kinetic versus interaction energy within the emulsion phase}

Following the same preparation protocol developed in ~\cite{Amico2018}, we investigate the behavior of the mean interaction, kinetic, and release energy of the fermionic atoms after some evolution at strong repulsion. We combine the precise determination of the interaction energy, spectroscopically quantified by measuring the instantaneous interaction shift $\Delta_+(t)$, with the study of the atomic cloud size by time-of-flight expansion performed in the presence or absence of interactions, thereby yielding the release and kinetic energy, respectively.

For each interaction strength and evolution time, the mean kinetic energy $E_{kin}(t)$ per atom is obtained by fitting the expanded density profile to a two-dimensional Gaussian envelope, from which we extract the radial and axial widths $\sigma_\bot (t)$ and $\sigma_x(t)$.
In particular, we proceed as it follows: at a certain magnetic field we prepare the interacting gas by means of the fast RF transfer at $\Delta_+(0)$, followed by the cleaning optical blast, and we wait for a variable evolution time $t$ in the trap.
Immediately before switching off the dipole potential, we apply a 150\,$\mu$s pulse at $\Delta_+(t)$, which selectively transfers the remaining unpaired state-3 atoms back into the weakly interacting state 2. 
We then let the gas expand for a variable time-of-flight, during which the state-2 atoms undergo ballistic expansion (see Fig.~\ref{FigEkin_TOF}), and acquire a spin-selective absorption image.  
 Owing to the reduced atomic signal, especially at strong coupling and long evolution times, we typically expand the clouds for a time-of-flight not exceeding 3\,ms. 
This time is comparable to the radial trap period, but still shorter than the axial one. For this reason, the mean kinetic energy of the atoms is determined by fitting the expansion of the radial width $\sigma_{\bot}(t)$: $E_{kin} = \frac{3 M}{2} (\sigma^2_{\bot}(t) - \sigma_{\bot0}^2)/t^2$, where $\sigma_{\bot0}$ is the in-situ radial width of the cloud.

\begin{figure}[!t]
\centering
\vspace*{1pt}
\includegraphics[width= 8.6 cm]{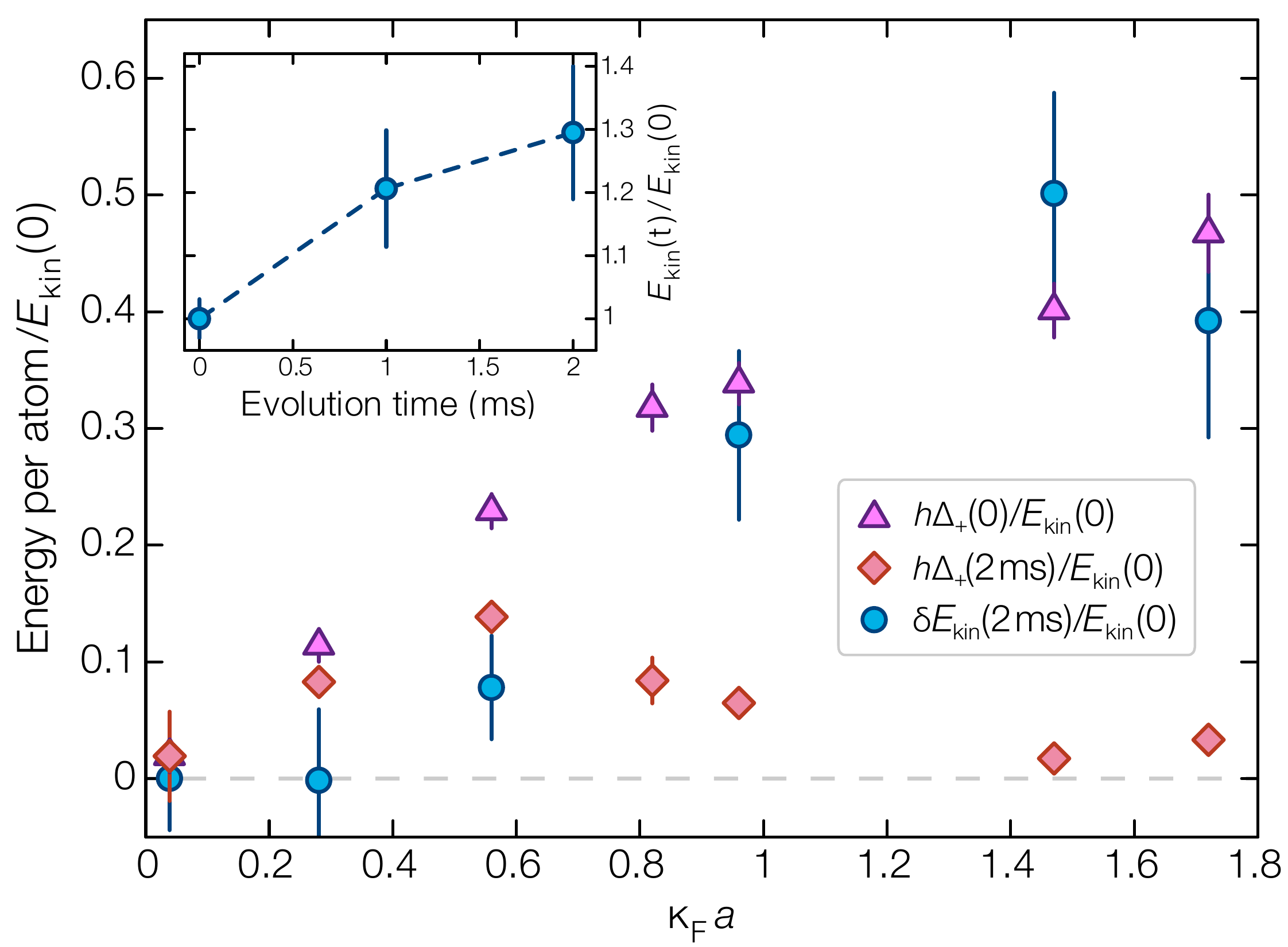}
\caption{\label{FigEkin}
Kinetic and interaction energy of state-3 atoms as a function of the interaction strength. 
Blue circles represent the increase of kinetic energy $\delta E_{kin} (2\,\textrm{ms})$ of the fermions measured via ballistic expansion after 2\,ms of evolution in the interacting state. For each interaction strength, $E_{kin}$  is obtained by averaging over typically five experimental realizations for each time of flight, which in turn is varied between 0.5\,and 3\,ms in 0.5\,ms steps. 
Purple triangles (red diamonds) denote the average interaction energy $\Delta_+(t)$ experienced by the atoms after zero ms (2\,ms) of evolution in the interacting state, obtained through RF spectroscopy. While $\delta E_{kin}(2\,\textrm{ms})$ and $\Delta_+(0)$ monotonically increase with the interaction strength, $\Delta_+(2\,\textrm{ms})$ exhibits a non-monotonic trend, with a maximum at about $\kappa_Fa \simeq 0.6$.
(Inset) Relative increase of kinetic energy of the fermions measured at different evolution times for $\kappa_Fa \simeq 1$. Only after some time,  the interaction energy is converted into kinetic one.
All data points are normalized to the mean kinetic energy $E_{kin}(0)$ measured for a non-interacting Fermi gas at $T/T_F = 0.12(2)$. 
}
\vspace*{0pt}
\end{figure}
In Fig.~\ref{FigEkin}, blue circles represent the trend of the relative increase of kinetic energy $\delta E_{kin}(2$\,ms$)= E_{kin}(2$\,ms$)-E_{kin}(0)$ normalized to the energy measured for a non-interacting sample at the same temperature $E_{kin}(0)$, recorded 2\,ms after the transfer  into the interacting state, as a function of the interaction strength. 
For initial interaction strengths $\kappa_Fa \gtrsim 0.6$, the gas kinetic energy sharply grows above its value at weak interactions, with a relative increase as large as 50$\%$. This trend is in good agreement with the results of previous studies \cite{Sanner2012}, and it is consistent with the development of anticorrelations between the two atomic spin states, which reduce the repulsive interaction energy at the price of an increased kinetic energy. The observed 50$\%$ increase in $E_{kin}$ at strong coupling is quantitatively consistent with the development of \textit{fully} spin-polarized domains, which are expected to cause a relative enhancement of $2^{2/3} \sim 1.6$ for homogeneous samples. On the other hand, such an increase could also arise from a combination of heating, associated with inelastic decay processes developing within the first milliseconds of evolution, and kinetic energy increase due to domain formation in a trapped system, which in turn would cause a factor $2^{1/3}\simeq 1.25$ enhancement of $E_{kin}$. 
In Fig. \ref{FigEkin}, we also display the interaction energies $h \Delta_+(0)$ (purple triangles) and $h \Delta_+(2\,\text{ms})$ (red diamonds), measured via RF spectroscopy at the beginning and after 2\,ms of evolution, respectively. 
While $h \Delta_+(0)$ monotonically increases with $\kappa_Fa$, $h \Delta_+(2\,\text{ms})$, after an initial growth, features a pronounced drop at strong couplings.

\begin{figure}[t!]
\centering
\includegraphics[width= 8.6 cm]{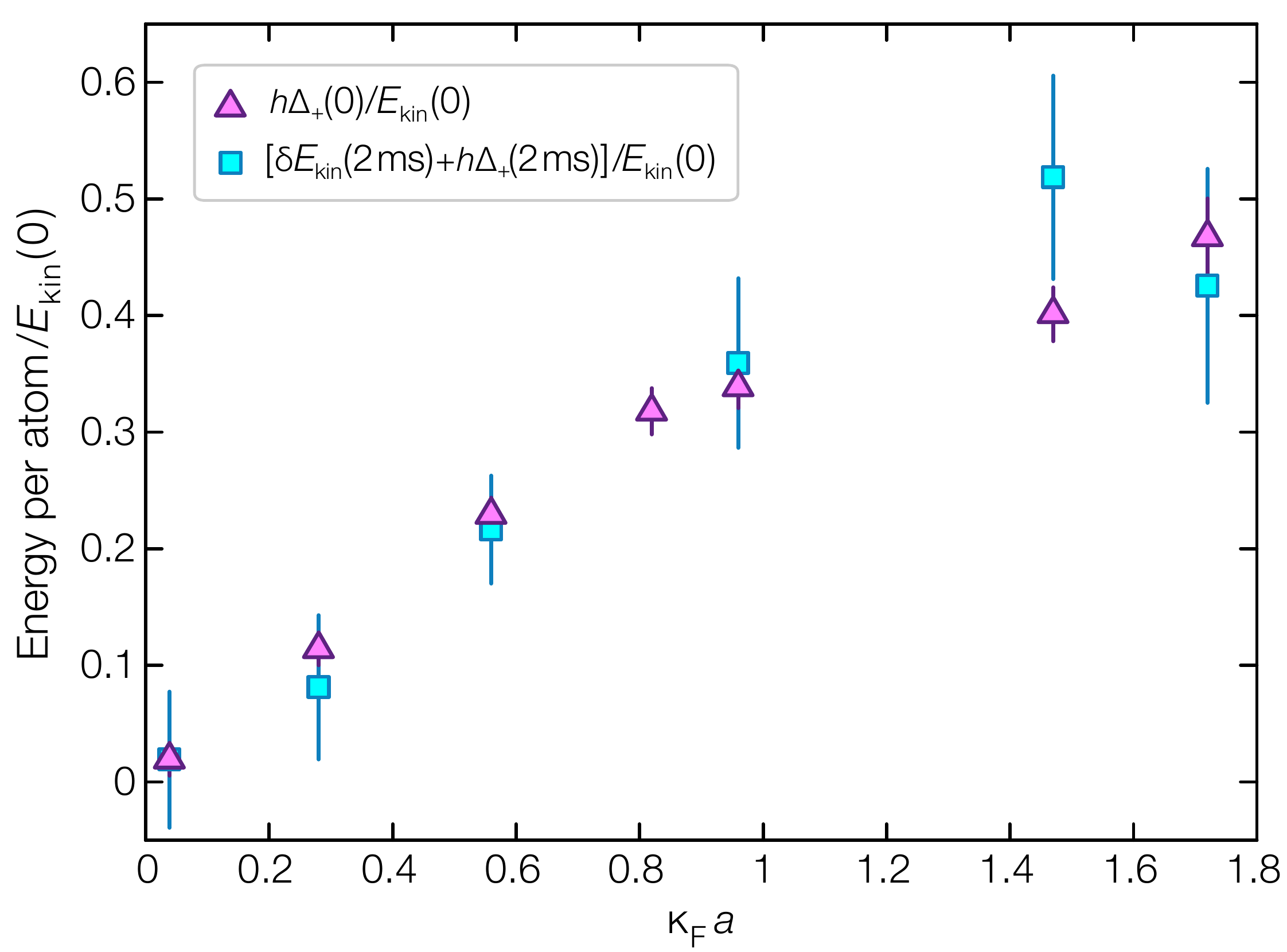}
\caption{\label{FigEkinEint}
Relative kinetic energy increase plus interaction energy of the state-3 atoms after 2\,ms of evolution in the trap (blue squares) and initial interaction energy (purple triangles) versus interaction strength. All data are normalized to the kinetic energy measured for a non-interacting Fermi gas at $T/T_F =0.12(2)$. The excellent overlap between the two quantities points to an almost perfect conversion of (initial) interaction into (final) kinetic energy, ruling out that the system heats up considerably over this timescales.  
\vspace*{0pt}}
\end{figure}

It is interesting to inspect how the variations of the mean interaction and kinetic energy of unpaired fermions, stemming from the evolution at strong interactions, compare with their initial interaction energy. 
In Fig.~\ref{FigEkinEint}, we compare the sum $\delta E_{kin}(2\,\text{ms}) + h \Delta_+(2\,\text{ms})$ with the initial interaction energy $h \Delta_+(0)$ at varying $\kappa_Fa$. 
The almost perfect overlap between the two quantities shows that, at least within such a timescale of about $\sim 100 \tau_F$, heating effects are insignificant, the interaction energy drop being fully converted into kinetic energy. 
This observation consistently points to the development of polarized domains as predicted by the Stoner mechanism, but in coexistence with a  sizable molecular fraction.

\subsection{Release versus kinetic energy measurements}

\begin{figure}[t!]
\centering
\vspace*{0pt}
\includegraphics[width= 8.6 cm]{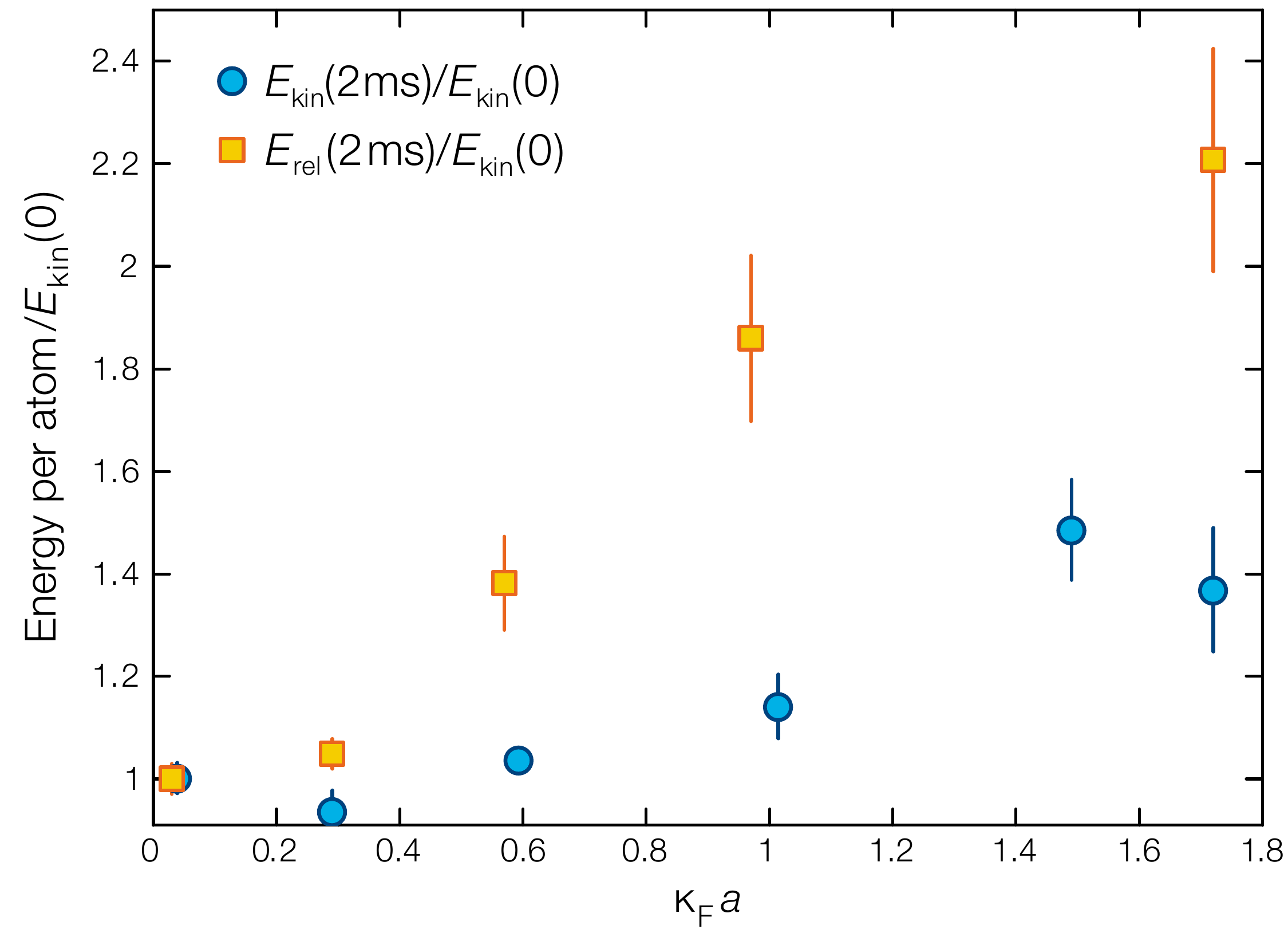}
\caption{\label{FigERel}
Kinetic (blue circles) and release energy (yellow squares) of the state-3 atoms after 2\,ms of evolution in the trap at varying interaction strength. All data are normalized to the kinetic energy measured for a non-interacting Fermi gas at $T/T_F=0.12(2)$. For all interaction regimes explored, the release energy always exceeds the kinetic one, owing to collisional hydrodynamic expansion of the atom-molecule mixture during the time-of-flight expansion, which causes an energy transfer from the molecular to the fermionic cloud.  
}
\end{figure}

Our observation of an almost perfect conversion of $h \Delta_+(0)$ into $E_{kin}(t)$ at strong repulsion appears inconsistent with the characterization reported in the Supplementary Materials of Ref. \cite{Sanner2012} of the release energy $E_{rel}$, which was found to greatly exceed $E_{kin}$ for all interaction regimes.
This apparently different behavior can be first attributed to the fact that in Ref.~\cite{Sanner2012} the system was presumably not allowed to evolve for enough time in the repulsive state. Indeed, as shown in the inset of Fig.~\ref{FigEkin} for $\kappa_Fa \simeq 1$, the kinetic energy progressively increases with the time spent in the interacting regime, reaching a steady value for times exceeding 2\,ms, on the order of half of our radial trap period. 
We thus perform an independent measurement of the release energy after a 2\,ms-long evolution time of the 1-3 repulsive mixture. 
The data are recorded by letting the gas expand in the presence of interactions, and selectively transferring the state-3 atoms back in the state 2 only right before acquiring the image.
The results are presented in Fig.~\ref{FigERel}. There we compare, for different $\kappa_F a$ values, the previously determined kinetic energy $E_{kin} (2\,\textrm{ms})$ with the release energy $E_{rel} (2\,\textrm{ms})$, both normalized to the initial $E_{kin}(0)$ value. 

In spite of a longer time spent in the interacting state, also in this case $E_{rel}$ always exceeds $E_{kin}$, consistent with the observations of Ref.~\cite{Sanner2012} and seemingly in contrast with the trend reported in Fig.~\ref{FigEkinEint}. 
This mismatch is solved by noticing that the release energy provides a reasonable estimate of $E_{kin}+E_{int}$ only if the gas features a negligible molecular fraction, since bosonic dimers can store a significant amount of interaction energy. As a consequence, 
the measurement of $E_{rel}$ in the presence of a substantial paired population is crucially affected by the energy transfer between molecular and atomic samples during the collision-driven hydrodynamic expansion, clarifying the seemingly contrasting results from previous studies \cite{Jo2009,Sanner2012}.
Most importantly, these observations also reveal that the trend of the fermionic components in the semi-stationary state, reached after a few ms evolution at strong coupling, appears  surprisingly compatible with the Stoner's criterion, in spite of the presence of a sizable paired fraction, totally neglected in the original Stoner model.

\subsection{Characterization of polarized domains via Rabi oscillations}

To gain deeper insights into the breakdown of the homogeneous repulsive Fermi liquid towards the emulsion state, we perform Rabi oscillation measurements on the state-3 atoms. The underlying idea of this characterization is that particles arranged into polarized domains will substantially differ from the quasi-particles composing the paramagnetic Fermi liquid phase, as they will behave as bare, non-interacting atoms. As such, they will feature a unity quasi-particle weight \cite{Massignan2014}. 
For interacting fermions, a reduction of the quasi-particle residue leads to a renormalization of the Rabi frequency, as demonstrated by recent experiments in the polaronic limit \cite{Kohstall2012, Scazza2017}. 
By monitoring the Rabi oscillations,  both in the Fermi liquid and in the emulsion regimes, we therefore obtain spectroscopic information not only about energy shifts, but also on the coupling matrix elements and thus about the character of the quasi-particles. 

For this measurement we set the RF frequency at the previously determined $\Delta_+(t)$, while the RF power is adjusted to the maximum value allowed by our apparatus. 
We extract the frequency $\Omega$ and the damping rate $\gamma_R$ by fitting the data with the function $f(t) = A\,e^{-\Gamma_R t} + B\,e^{-\gamma_R t} \cos(\Omega t)$, describing a Rabi oscillation at frequency $\Omega/(2\pi)$ with a damping $\gamma_R$ and a state-3 atomic population decay rate $\Gamma_R$ (with $A, B \simeq 0.5$) \cite{Scazza2017}. 
\begin{figure}[t!]
\centering
\includegraphics[width= 8.5 cm]{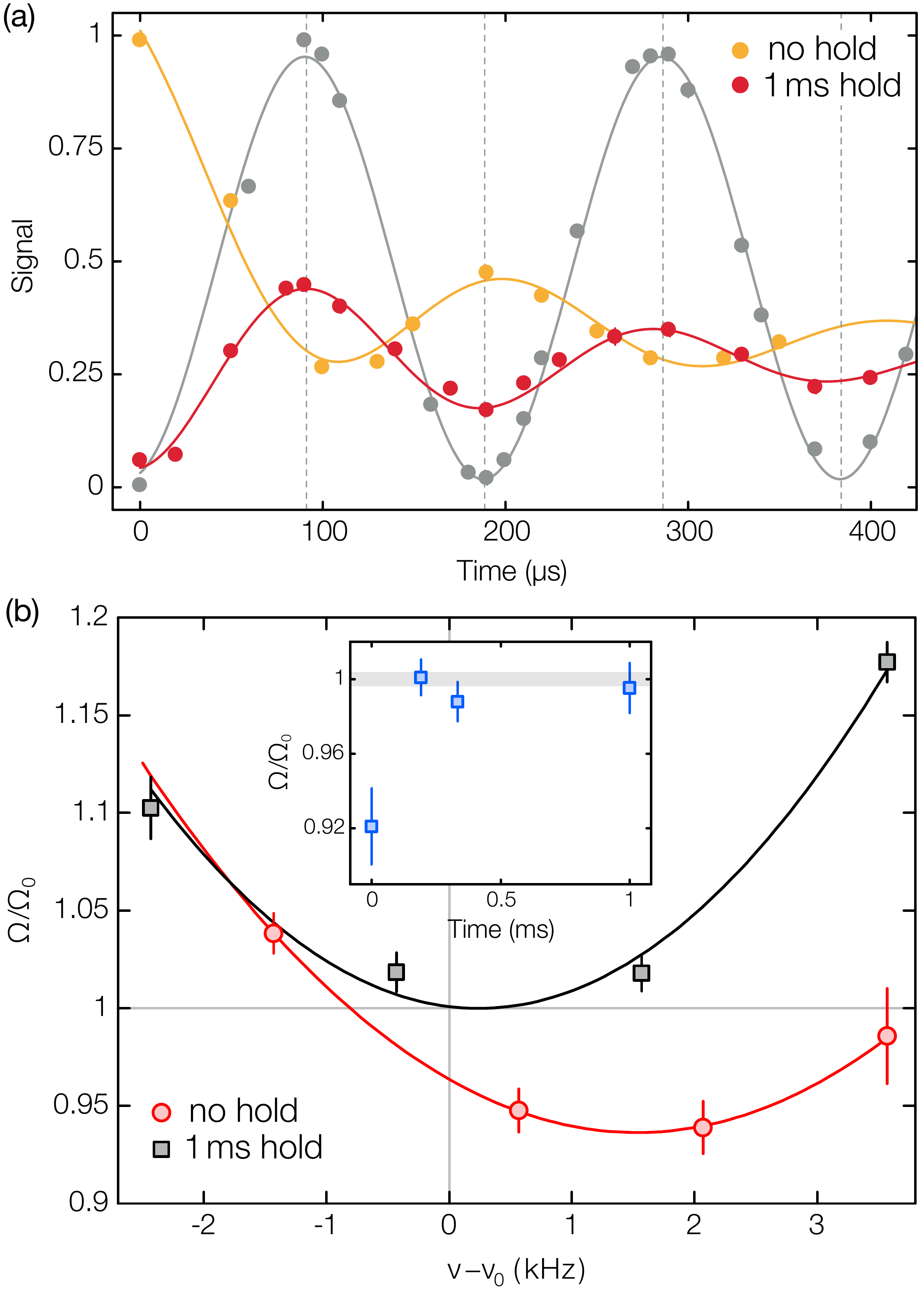}
\caption{\label{figRabi}
Characterization of the system by Rabi oscillations. (a) Typical Rabi oscillation measurements for the emulsion state (red circles), and the Fermi liquid system (yellow circles). The bare atom Rabi oscillation is also shown as a reference (grey circles).
(b) Behavior of the Rabi frequency versus detuning from the bare-atom resonance at $\kappa_Fa \simeq 2$, both for the emulsion phase (gray squares), and in the paramagnetic system (red circles). For reaching the emulsion state, we let the system evolve in the interacting state for 1\,ms, after which we drive Rabi oscillations setting the frequency around the location of the atomic peak. For the paramagnetic system, we drive Rabi oscillations by starting with a 1-2 weakly-interacting paramagnetic mixture. In both cases the RF power was set to the maximum value allowed by our amplifier, resulting in free-atom Rabi frequencies of about 5.2 kHz.
Atoms driven on the 2-3 transition in the mixed system feature a considerably strong renormalization of the frequency, $\Omega <\Omega_0$, which reaches its minimum at $\Delta_+(0) \sim 2\,$kHz. In turn, atoms in the emulsion case oscillate at the bare atom Rabi frequency, which is minimized for nearly zero detuning $\nu - \nu_0 \simeq 200$\,Hz.
}
\end{figure}
In the Fermi liquid regime, addressed by Rabi flipping state-2 atoms with the RF field set at $\Delta_+(0)$ (see yellow circles in Fig. \ref{figRabi}a), 
the renormalization of quasiparticle coherence,  
encoded in a quasiparticle weight smaller than unity, results in $\Omega/\Omega_0 \leq 1$ \cite{Scazza2017, Kohstall2012}. 
Here, $\Omega_0$ denotes the free-atom Rabi frequency, experimentally calibrated using a spin-polarized state-2 Fermi gas (gray circles in Fig. \ref{figRabi}a).
In turn, in the emulsion phase, emerging in our system shortly after a first pump pulse and resulting in $\Delta_+(t) \simeq 0$, coherent oscillations occur at a Rabi frequency compatible with that of a spin-polarized gas, $\Omega/\Omega_0=1$, yet featuring a finite damping (red circles in Fig. \ref{figRabi}a). 
This trend of the Rabi frequency is summarized in Fig.~\ref{figRabi}b for $\kappa_Fa \simeq 2$, where we compare $\Omega/\Omega_0$ extracted by monitoring the $2 \leftrightarrow 3$ Rabi flopping at different detunings $\Delta$ of the RF drive from the non-interacting transition frequency and two different hold times preceding the Rabi driving. 

In both cases, the Rabi frequency quantitatively matches the usual relation $\Omega_{\Delta} \sim \sqrt{\Omega^2+\Delta^2}$ expected by coherently coupling two discrete energy levels. One should also note that in the zero (1\,ms) hold time the minimum value $\Omega/\Omega_0<1$ ($\Omega/\Omega_0=1$) is reached at a positive (zero) detuning from the free atom transition resonance.
This observation further supports the emergence of the emulsion state at strong coupling: in such a state spin-polarized atomic clusters may form, within which state-3 fermions behave as non-interacting particles with unity quasiparticle weight.  
On the other hand, the damping of such coherent oscillations remains sizable, see Fig. \ref{figRabi}a. This agrees with the picture of micro-domains of only a few interparticle spacings, so that the heterogeneous spin texture may rearrange itself during the Rabi cycles, allowing for collisional decoherence effects to develop. 

The quick development of short-range anti-correlations and micro-sized domains \cite{Amico2018} is also signaled by the rapid variation of $\Omega/\Omega_0$ with the time spent in the strongly repulsive regime. As shown in the inset of Fig.~\ref{figRabi}b, the measured Rabi frequency matches  the bare atomic one already  about 10\,$\tau_F$ after the transfer into the interacting regime.
This conclusion is also confirmed by the absence of any signature of macroscopic phase separation, discussed in our previous works \cite{Amico2018,Jo2009,Sanner2012}.

\section{Stability of the quantum emulsion}
%
The measurements presented so far reveal the highly correlated character of the emulsion state. While a large fraction of the two initial spin  components is rapidly converted into atom pairs, the surviving fermions exhibit aspects of ferromagnetic behavior predicted for a purely repulsive Fermi system. In this section we move to address two key questions: The first one concerns the trend of the system at longer times, and the second one is about the role of the initial degree of degeneracy. On the one hand, a strongly correlated state should not to exist at high temperatures.  On the other hand, one would expect a sizable heating associated with the energy released in the pair formation processes, of about $2 \epsilon_F$ at $\kappa_F a \simeq 1$.

\subsection{Stabilization and melting of the quantum emulsion}

We now focus on measurements targeted to test the metastability of this exotic many-body system. Specifically, we investigate the response of the emulsion state to slow changes of the interaction strength, by monitoring the interaction shift $\Delta_+$ and the atomic peak area, revealed by a probe $3 \rightarrow 2$ spectroscopy pulse. 
\begin{figure}[t]
	\centering
	\includegraphics[width= 8.6 cm]{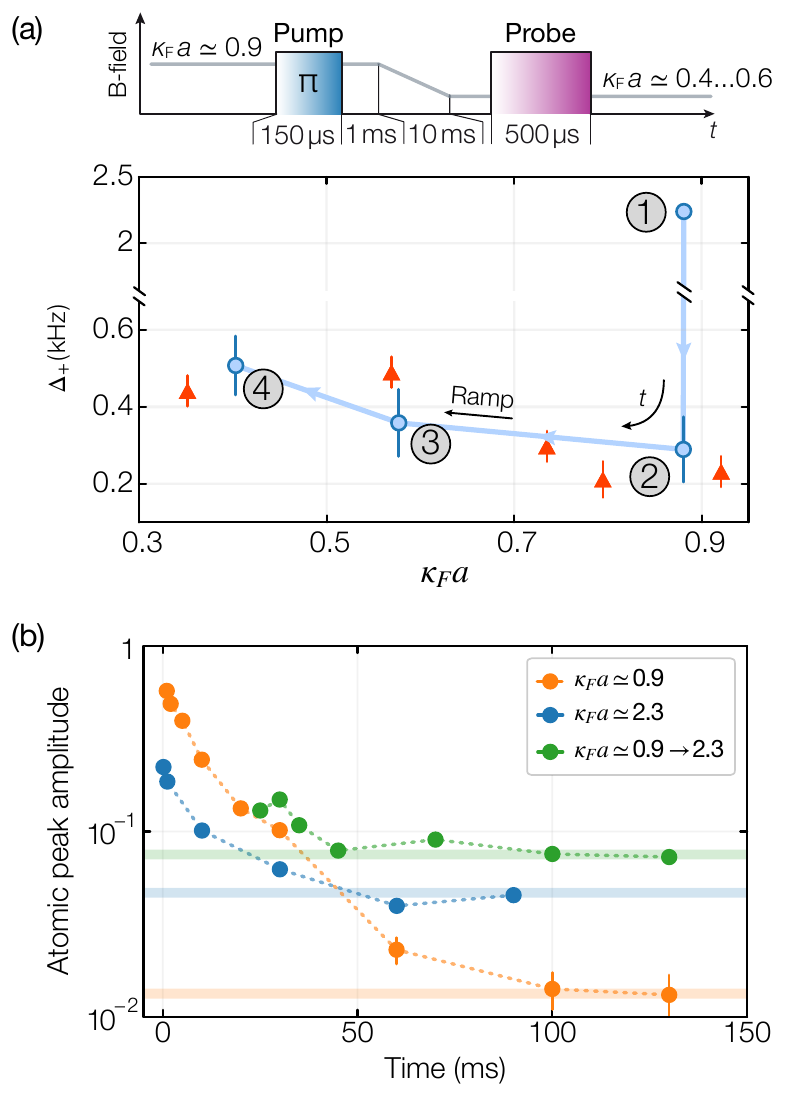}
		\caption{Melting and metastability of the emulsion phase.  (a) After rapidly accessing the emulsion regime at $\kappa_F a\simeq 0.9$, a  magnetic field ramp to lower $(\kappa_F a)_\text{f}$ values melts any inhomogeneous structure, re-establishing an interaction shift (blue circles) compatible with $\Delta_+(\infty)$ obtained for a gas initialized at $(\kappa_F a)_\text{f}$ (red triangles). (b) Using a similar protocol as in (a), but ramping towards stronger interactions, pairing is substantially suppressed and the emulsion is stabilized (green). Atom loss dynamics measured at fixed $\kappa_F a\simeq 0.9$ (orange) and $\kappa_F a \simeq 2.3$ (blue) are shown for comparison. Error bars in both figures represent the statistical error of the fits employed to extract the center and amplitude of the atom spectral peak, respectively.}
\label{fig_MeltingLifetime}
	\vspace*{1pt}
\end{figure}

In Fig.~\ref{fig_MeltingLifetime}a, we present a measurement of $\Delta_+$ obtained by letting first the emulsion develop for 1\,ms at $\kappa_F a \simeq 0.9$ (steps 1 and 2), and by successively decreasing the interaction to lower final $(\kappa_F a)_\text{f}$ values (marked as 3 and 4 in the figure) through a linear magnetic field ramp. 
The interaction shift $\Delta_+$, measured after the Feshbach sweep, progressively increases from its smaller value at $\kappa_F a \simeq 0.9$, matching within experimental uncertainty the asymptotic shift $\Delta_+(\infty)$ recorded for a system directly initialized at $(\kappa_F a)_\text{f}$.
The restoration of the interactions shifts at lower $(\kappa_F a)_\text{f}$ values that correspond to the paramagnetic Fermi liquid phase, observed in spite of a partial (irreversible) atom-molecule conversion, is consistent with an interaction-driven melting of micro-domains within the heterogeneous phase. Moreover, such observation further rules out that the negligibly small $\Delta_+$ values obtained at sufficiently long evolution times in the strong coupling regime \cite{Amico2018} arise as a trivial consequence of the heating associated with the pairing processes. If heating, rather than anti-correlations, caused the drop in the  interaction energy, a sizable repulsive shift could not be restored by decreasing $\kappa_F a$, and it would be inconsistent with the observed reversibility.

Another non-trivial feature of the peculiar atom-molecule mixture reached at long time  is highlighted in Fig.~\ref{fig_MeltingLifetime}b. Here, the green circles display the evolution of the atomic peak amplitude obtained by first creating the emulsion at $\kappa_F a \simeq 0.9$, and successively increasing the coupling to $(\kappa_F a)_\text{f} \simeq 2.3$. This trend is contrasted with the behavior of a system evolving at the initial (blue circles) and final (orange circles) interaction strengths, respectively. Let us first focus on the dynamics featured by the gas at the two fixed interaction regimes. Differently from the monotonic increase of the pairing rate observed right after the transfer into the repulsive 1-3 state \cite{Amico2018}, the remaining atomic population at long evolution times greatly increases with $(\kappa_F a)$. Such a trend could be in principle interpreted as the establishment of chemical equilibrium between atoms and pairs in a hot, incoherent paramagnetic state. For stronger couplings, the molecular binding weakens and, as a consequence, the surviving atomic fraction at fixed temperature is expected to increase \cite{Chin2004}. 
However, this interpretation is inconsistent with the trend of the fermionic fraction revealed by letting the system initially evolve at $\kappa_F a \simeq 0.9$, and then \textit{increasing} the interaction strength. Since in this case, the measured temperature is within experimental uncertainty the same as the one measured at the final field without magnetic sweep, the significantly larger atomic fraction appears irreconcilable with a chemically-equilibrated mixture. Rather, the increased coupling, externally imparted after the emulsion is created at $\kappa_F a \simeq 0.9$, strongly inhibits inelastic recombination processes. This behavior can be naturally explained if we assume that the atomic density distributions are spatially inhomogeneous: In that case, pairing can only occur if atoms from oppositely-polarized clusters diffuse throughout intermediate mixed or molecular layers. The latter act as repulsive barriers monotonically increasing with $\kappa_F a$, hence suppressing pair formation at strong coupling. In addition to providing further signature of micro-scale phase segregation, Figure~\ref{fig_MeltingLifetime}b data also highlight the extraordinary stability of such a highly-correlated state of matter: Over timescales exceeding 100 ms, i.e.\,approximately $10^4$ $\tau_F$, comparable to the lifetime of molecular superfluids \cite{Varenna2007}, little dynamics occurs, mainly ascribed to plain evaporation from the dipole trap.

\subsection{Temperature dependence of the emulsion regime}

In this section we explore the effect of an initially lower degree of degeneracy on the stability of the emulsion phase.
For this purpose, we determine the center and the area of the atomic peak from probe $3 \rightarrow 2$ spectra, recorded by letting either a highly degenerate sample at $T/T_F=0.12(2)$, or an initially hotter gas at $T/T_F\simeq 0.3$, evolve in the interacting state. 
\begin{figure}[t]
	\centering
	\includegraphics[width= 7.6 cm]{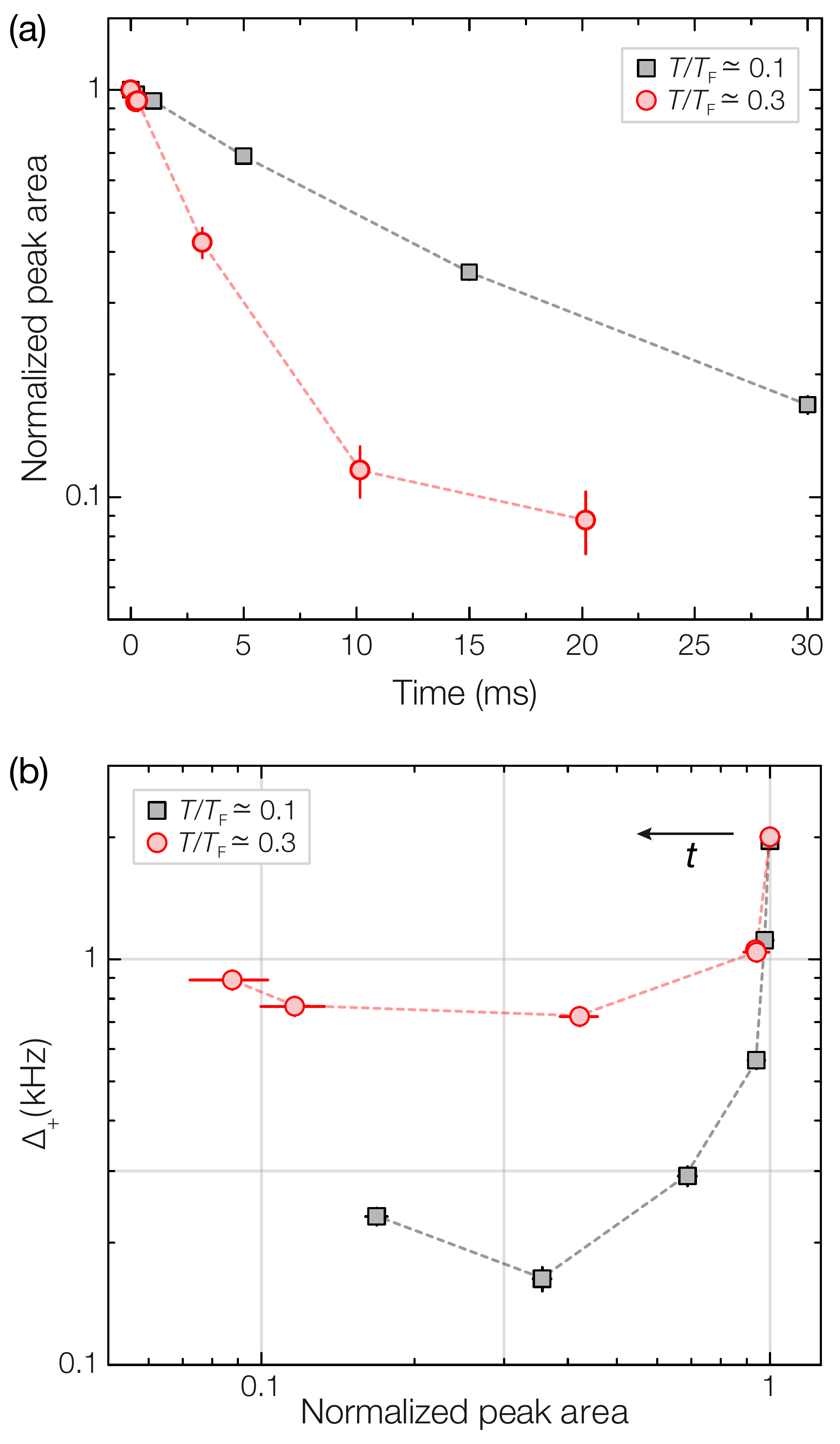}
		\caption{Temperature dependence of the emulsion phase. (a) Evolution of the normalized atomic peak area, directly yielding the surviving fermion fraction, recorded at $\kappa_F a \simeq 1.2$ for highly (gray squares) and moderately (red cyrcles) degenerate mixtures, see legend. (b) Interaction shift experienced by the surviving fermions as a function of their relative population for the two temperature regimes detailed in the legend. The stronger atom loss and the larger residual interaction shifts revealed for the less degenerate sample point to the low-temperature nature of the emulsion state, inconsistent with a paramagnetic atom-molecule mixture at chemical equilibrium.}
\label{RFHot}
	\vspace*{1pt}
\end{figure}
For both temperature regimes investigated, pump and probe RF pulses are fixed to be 160 and 300\,$\mu$s, respectively.
Examples of this characterization are presented in Fig.~\ref{RFHot}. For all data sets, we adjust the trap frequencies and the atom number so as to keep the same central density, hence Fermi energy, within a 20$\%$ uncertainty. 
This results in similar initial $\Delta_+(0)$ values for similar values of $\kappa_F a \sim 1.2$.
From Fig.~\ref{RFHot}a, one can notice how a higher $T/T_F= 0.3$ yields a faster atom-to-molecule conversion, relative to the one detected at $T/T_F=0.12(2)$. Moreover, the semi-stationary fermionic fraction reached at long evolution times is significantly larger for highly degenerate samples, rather than for the hotter ones.  
This is opposite to what one would expect for an atom-molecule mixture in thermal and chemical equilibrium, where at $T=0$ the sample is purely molecular, and where a higher temperature yields an increased fraction of unpaired atoms \cite{Chin2004}.
  
 As the evolution of the interaction energy is concerned, see  Fig.~\ref{RFHot}b, our measurements reveal that the drop of the interaction shift at fixed atomic fraction becomes less pronounced for a decreased gas degeneracy, remaining sizable even once semi-stationary conditions have been reached.
This trend suggests that the emulsion phase is attained only in the highly degenerate regime, at least for $\kappa_Fa \sim 1$. Namely, lowering the temperature leads to smaller interaction energies. This is similar to the expectation of the Stoner model, where only at sufficiently low temperatures the gas becomes ferromagnetic due to macroscopic phase separation. Here, we observe this behavior within a micro-emulsion state.

The data in Figure~\ref{RFHot}b also rule out that the $\Delta_+(t)$ drop is affected by higher order partial wave contributions connected with the atom-molecule scattering. In particular, $p-$wave atom-dimer interactions are known to be attractive \cite{Jag2014} also for equal mass mixtures, and they are expected to increase, relative to the $s-$wave ones, for increased collision energy \cite{Levinsen2011}. If these were sizable, the overall drop of $\Delta_+$ should become more pronounced for higher temperature regimes. This scenario appears inconsistent with our observation in Figure~\ref{RFHot}b, where the most significant reduction of the interaction shift is attained in the highly degenerate regime.

\section{Probing micro-scale inhomogeneity}

The characterization of the semi-stationary state reached at long evolution times at strong coupling is complicated since micro-scale inhomogeneity cannot be directly detected through spin-selective \textit{in situ} imaging of the cloud. This is due to line-of-sight integration of our three dimensional samples and to the small estimated domain size, only slightly larger than our imaging resolution of about $1.3\mu$m.
In this section we discuss two additional characterizations that overcome this issue and that further point to the existence of a heterogeneous emulsion phase within the central, high density region of the cloud. These studies rely both on the local measurements of spin density fluctuations \cite{Recati2010, Sanner2011, Sanner2012, Meineke2012} and their spatial correlations \cite{Roscilde2007}, and on a special RF spectroscopy protocol involving a three spin state combination, which exploits a small state-2 non interacting component as a local probe for spin density inhomogeneities of the 1-3 mixture. 

\subsection{Noise correlation measurements}
Spin density fluctuations \cite{Recati2010, Sanner2011, Sanner2012, Meineke2012} and their spatial correlations \cite{Roscilde2007} are sensitive to the presence of spin inhomogeneities. In particular, the presence of spin-polarized clusters in a spin-balanced sample leads to an increased variance of the spin number fluctuations inside any probe volume with a transverse size comparable or larger than the imaging resolution \cite{Sanner2012} that, otherwise, smears out and thus increases the effective probe volume. In the simplistic case of Poissonian fluctuations, the enhancement of spin variance in a heterogeneous sample, relative to that of a paramagnetic one, would directly reflect the mean number of spins per domain \cite{Sanner2012}. Moreover, in an emulsion of spin domains with mean size $d$, spin density-density correlations at distance $R<d$ ($R>d$) are expected to be enhanced (reduced) relative to a homogeneous configuration \cite{Roscilde2007, Teubner1987}. 
Investigating these observables in our system thus provides valuable information about possible spatial structures at the micro-scale.
 %
\begin{figure}
\centering
\includegraphics[width=8.5cm]{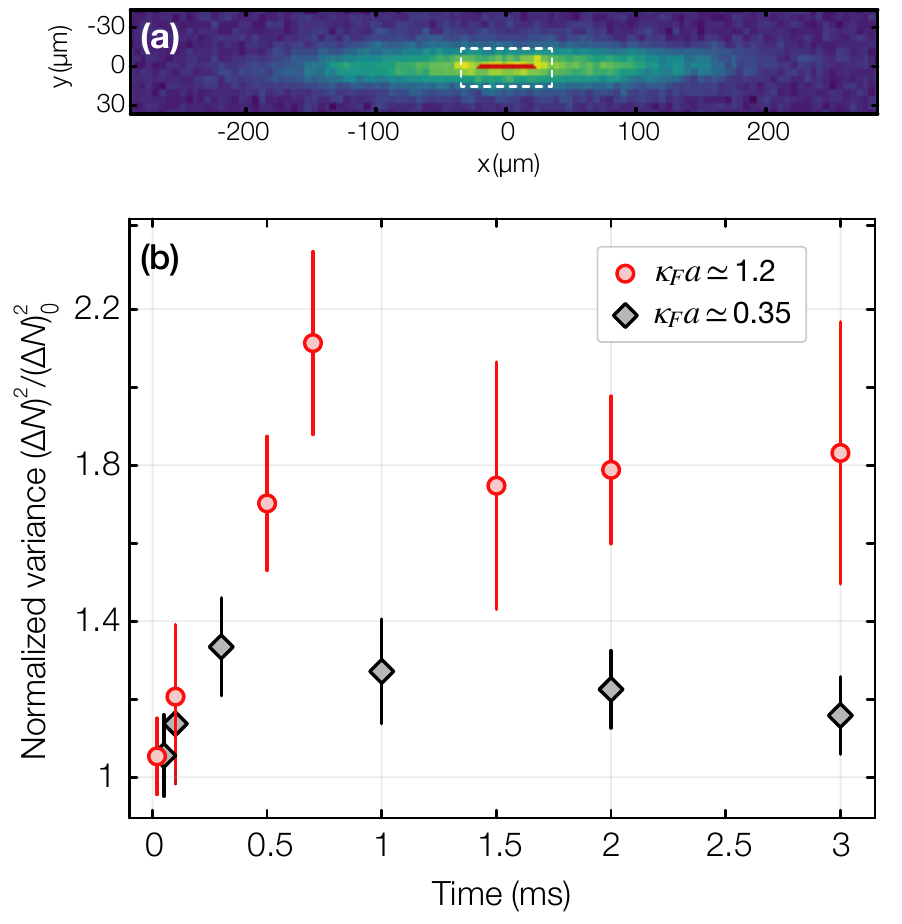}
\caption{Real time dynamics of atom number variance. (a) Example of an \textit{in situ} image of the state-3 component, employed to extract local information on spin density fluctuations and their spatial correlations. The region  $\Re$ around the cloud center that we selected for the analysis is marked by the dashed white rectangle.
(b) Atom number variance $(\Delta N)^2$ 
normalized to that of an ideal Fermi gas $(\Delta N)_0^2$, measured within the region $\Re$ at various evolution times at $\kappa_F a \simeq 1.2$ (red circles) and 0.35 (gray diamonds), respectively. 
Error bars denote the standard deviation of the mean.
\vspace*{1pt}}
\label{fig_var_center}
\end{figure}

Experimentally, density correlations are retrieved at different evolution times and interaction strengths by acquiring $\mathcal{N}  \approx 80$ \textit{in situ} absorption images of state-3 clouds. Owing to the shallow nature of the dimers, the imaging light probes equally atoms and molecules. 
We then extract both the mean local column-integrated density $\langle n(\textbf{r}) \rangle$, as well as the spin density-density correlation function, defined as $C(\textbf{r}, \textbf{r} + \textbf{R})= \langle \left( n(\textbf{r}) - \langle n(\textbf{r}) \rangle \right) \left(n(\textbf{r}+\textbf{R}) - \langle n(\textbf{r} + \textbf{R}) \rangle \right)\rangle$, where $\left\langle \dots \right\rangle$ denotes the mean over $\mathcal{N}$ experimental realizations. Focusing within a central, denser region $\Re$ of the gas (of $60 \cdot 30 \,\mu$m$^2$ and indicated as a white rectangle in Fig. \ref{fig_var_center}a), we then evaluate the mean azimuthally-averaged correlator $C(R)$ at distance $R$ therein:  
$C(R)=\left\langle (N_r-\overline{N}_r)(N_{r+R}-\overline{N}_{r+R}) \right\rangle_{\Re}$. Here  
$\overline{N}_{r}$ stems for the average state-3 number at position $r$ obtained from all $\mathcal{N}$ images, and $\left\langle \cdot \right\rangle_{\Re}$ is an average over all pixels within the region $\Re$. 
From $C(R)$, one can also straightforwardly obtain the state-3 number variance $(\Delta N)^2 \equiv C(0)$.
%

Figure~\ref{fig_var_center}b presents the evolution of $(\Delta N)^2$ normalized to that of a non-interacting gas $(\Delta N)_0^2$, for $\kappa_F a \simeq 0.35$ and 1.2, respectively. 
While for the weaker interaction we observe only a small increase of fluctuations, compatible with a moderately enhanced spin susceptibility \cite{Pilati2010, Recati2010}, for $\kappa_F a \simeq 1.2$ one can notice a more pronounced growth of $(\Delta N)^2$. The timescale of such evolution appears consistent with that of the spectral response at the same $\kappa_F a$ reported in our previous study Ref. \cite{Amico2018}, suggesting a common microscopic origin for the dynamics of the two observables. 
The observed maximum two-fold increase of the column-integrated spin fluctuations $(\Delta N)^2$ rules out the formation of any larger domain, but it is consistent with spin separation occurring only on a small $\mu$m-scale of few interparticle spacings. This is in agreement with a domain-size estimate based on our recent measurement of the growth rate $\Gamma_{\Delta}$ for ferromagnetic anti-correlations \cite{Amico2018}, i.e. $\xi \approx 2\pi/\kappa_F$ \cite{Pekker2011}.

\begin{figure}[t]
\centering
\includegraphics[width=8.5cm]{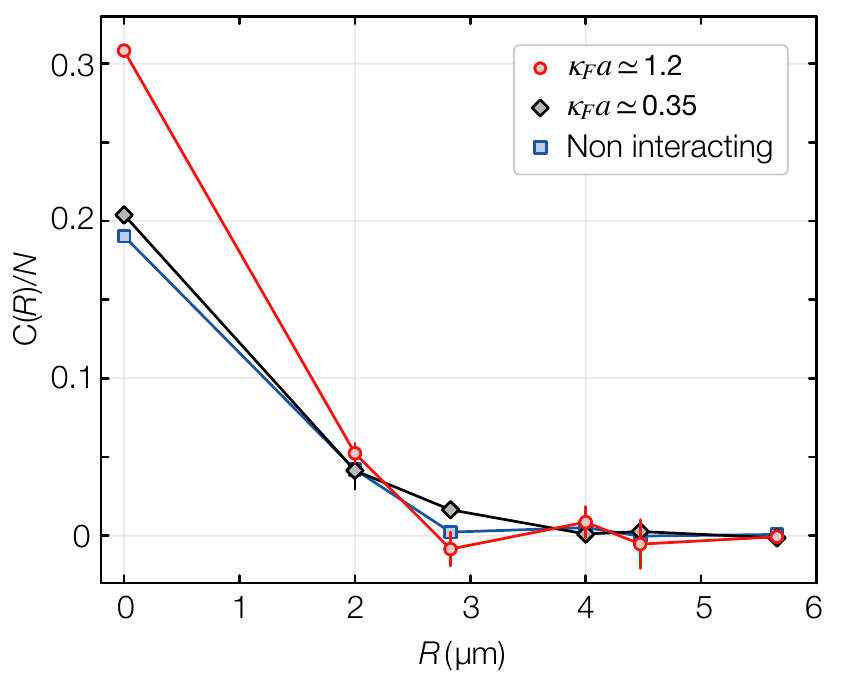}
\caption{Spatial correlations of spin density fluctuations. Correlation function of density fluctuations $C(R)$ normalized to the mean number per probe volume $N$, measured 700 $\mu$s after a quench to $\kappa_F a\simeq 1.2$ (red circles) and 0.35 (gray diamonds).
$C_0(R)$ of a weakly-interacting gas (blue squares), reflecting the point spread function of our imaging system, is shown for comparison.
Error bars denote the standard deviation of the mean.
%
\vspace*{1pt}}
\label{FIG_CorrFunc}
\end{figure}
The absence of larger domains can also be inferred from the extracted trend of $C(R)$. Fig.~\ref{FIG_CorrFunc} shows $C(R)$ normalized to the average atom number $N$ per probe volume, recorded after 700 $\mu$s of evolution at $\kappa_F a \simeq 0.35$ and 1.2, respectively. For the former case (gray diamonds), $C(R)$ features a trend similar to that measured in weakly-interacting samples, associated with the point spread function of our imaging system. 
For $\kappa_F a \simeq1.2$, aside the aforementioned enhancement of C(0) with respect to the weakly-repulsive gas, $C(R)$ features a sharper drop for $R \leq 2.5\,\mu$m, and it eventually turns slightly negative around $R\sim 3\mu$m. 
Such a trend, qualitatively matching the one expected for a micro-emulsion heterogeneous phase \cite{Roscilde2007, Teubner1987}, implies a correlation length comparable with, or smaller than our optical resolution.
It is also important to remark that our results are fully consistent with the ones previously reported in Ref.~\cite{Sanner2012}, although these latter observations were obtained by using slightly different experimental procedures.
\subsection{RF spectroscopy on a three-component mixture}

\begin{figure}[t!]
\vspace*{0pt}
\centering
\includegraphics[width= 8.6 cm]{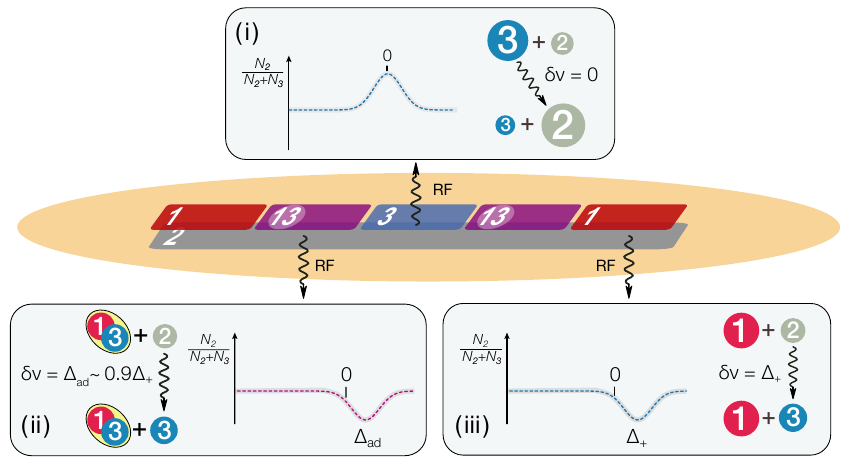}
\caption{
Schematic illustration of the probe spectroscopy on a three-component $1-2-3$ mixture. An RF pulse near resonant with the bare $2 \leftrightarrow 3$ atomic transition may cause different transitions when a weakly interacting state-2 Fermi gas is immersed in the emulsion phase. (i) For zero detuning, it transfers state-3 atoms arranged within polarized clusters back in the weakly interacting state 2, yielding a positive peak in the spectroscopy signal $N_2/(N_2+N_3)$. (ii)-(iii) For positive detunings, state-2 atoms, that are spread all over the density distributions of the other atomic and molecular components, can be transferred into the resonant 3 state. Irrespective of whether the state-2 fermions locally experience a molecular or a state-3 atomic bath, the optimum transfer occurs at positive detunings $\Delta_+ \simeq \Delta_{ad}$, and it yields a negative peak in the spectral response.}
\label{2peakstheo}
\end{figure}

Another way to probe micro-scale phase separation within the emulsion state is enabled by a slightly modified pump-probe spectroscopy protocol: In contrast with the one employed for all studies discussed throughout this paper and in our previous work Ref. \cite{Amico2018}, for this measurement
the optical blast following the RF pump pulse is not applied. 
In this case, the gas  comprises three different spin states and, after some evolution time, four different components, see sketch in Fig. \ref{2peakstheo}: unpaired state-1 and -3 resonantly interacting fermions, $1-3$ molecules created by pairing processes during the dynamics, and a small fraction of state-2 atoms, interacting only weakly with all other kinds of trapped particles. 
In general, the presence of three distinguishable fermionic states has detrimental effects on the stability of the trapped gas, limiting the lifetime of the overall population in the trap through three-body inelastic decay processes towards deeply-bound molecular states \cite{Ottenstein2008, Huckans2009}. On the other hand, we have checked that a small state-2 population not exceeding 30$\%$ of that of state-3, does not cause substantially enhanced losses within the first 2\,ms of evolution. 
As such, state-2 fermions can be considered as weakly-interacting spectators of the correlated many-body dynamics in the $1-3$ mixture and the emergent emulsion phase. In particular, while state-1 and -3 atoms, as well as $1-3$ pairs, may arrange within spatially distinct micro-domains, as depicted in the skecth in Fig. \ref{2peakstheo}, the state-2 component occupies the whole trap volume, featuring a density distribution of a weakly-interacting Fermi gas. 

%
Application of a $3 \leftrightarrow 2$ spectroscopy pulse after some evolution time within such a four component atom-pair admixture is expected to induce different types of transfer processes, schematically identified in Fig. \ref{2peakstheo}: (i) At a detuning $\Delta_+(t) \simeq 0$ from the bare $2\leftrightarrow 3$ transition, it can transfer a state-3 fermion within a polarized domain composing the emulsion phase, back into the weakly interacting spin state 2. 
On the contrary, an atom initially in state 2 experiences a full overlap with all other mixture components, in particular with leftover state-1 fermions and with $1-3$ pairs. As a consequence, the resonance RF frequency enabling its transfer into the interacting state will be blue shifted, relative to the bare atomic transition. Since the overall density is weakly affected by the development of the emulsion state, and since atom-dimer interactions at low temperature are essentially as repulsive as the atom-atom ones, the associated spectroscopy signal will result in a negative peak centered at positive detunings, essentially indistinguishable one from another: (ii) $\Delta_{ad}\simeq 0.9\Delta_+(0)$ if the state-2 fermion is overlapped with a molecular cluster, or (iii) $\Delta_+(0)$ when surrounded by unpaired state-1 atoms.   
Finally, for large and negative detunings, the probe pulse may also drive bound-to-free dissociation processes of the kind $1-3 \mapsto 1+2$, or vice versa free-to-bound $1+2 \mapsto 1-3$ transitions, not shown in the sketch. As a result of these different possible transfers (and neglecting pair dissociation or association processes), the overall spectral response within a spatially heterogeneous phase is expected to exhibit a characteristic peak-plus-dip structure.  

A totally different scenario would result for a paramagnetic phase. In that case, all components would feature full spatial overlap throughout the trap: Hence, spin-injection processes of the kind (i) would yield a positive peak centered at the same positive detunings that characterize transitions of the kind (ii) and (iii), respectively. In the paramagnetic case, the overall spectral response would thus feature a single peak structure on top of a non-zero background signal, whose total amplitude (positive or negative) solely depends upon the ratio between the state-2 and state-3 populations, initially adjusted by the pump pulse.  
The presence or absence of a heterogeneous phase within our trapped sample can therefore be revealed by the spectral response of a three-state sample. 

We apply this conceptual strategy in the experiment by initializing the system at strong repulsion, $\kappa_Fa \sim 1.7$, where the initial interaction shift $\Delta_+(0)$ is substantial, and the possible presence of a peak-plus-dip structure may be resolved in spite of collisional broadening of the spectral lines. 
In particular, starting from a balanced 1-2 mixture, we adjust the pump pulse parameters to create a three state combination with a leftover state-2  fraction approximately equal to 30\% of the initial population. We then allow the system to evolve for some variable time in the interacting state, eventually reaching the emulsion phase, before applying the second spectroscopy pulse.
From spin-selective absorption imaging of both state-2 and state-3 atoms recorded \textit{in situ} right after the probe pulse, we then determine the local spectral response, both within the central and the outermost region of the trapped sample.
\begin{figure}[t!]
\vspace*{0pt}
\centering
\includegraphics[width= 8.6 cm]{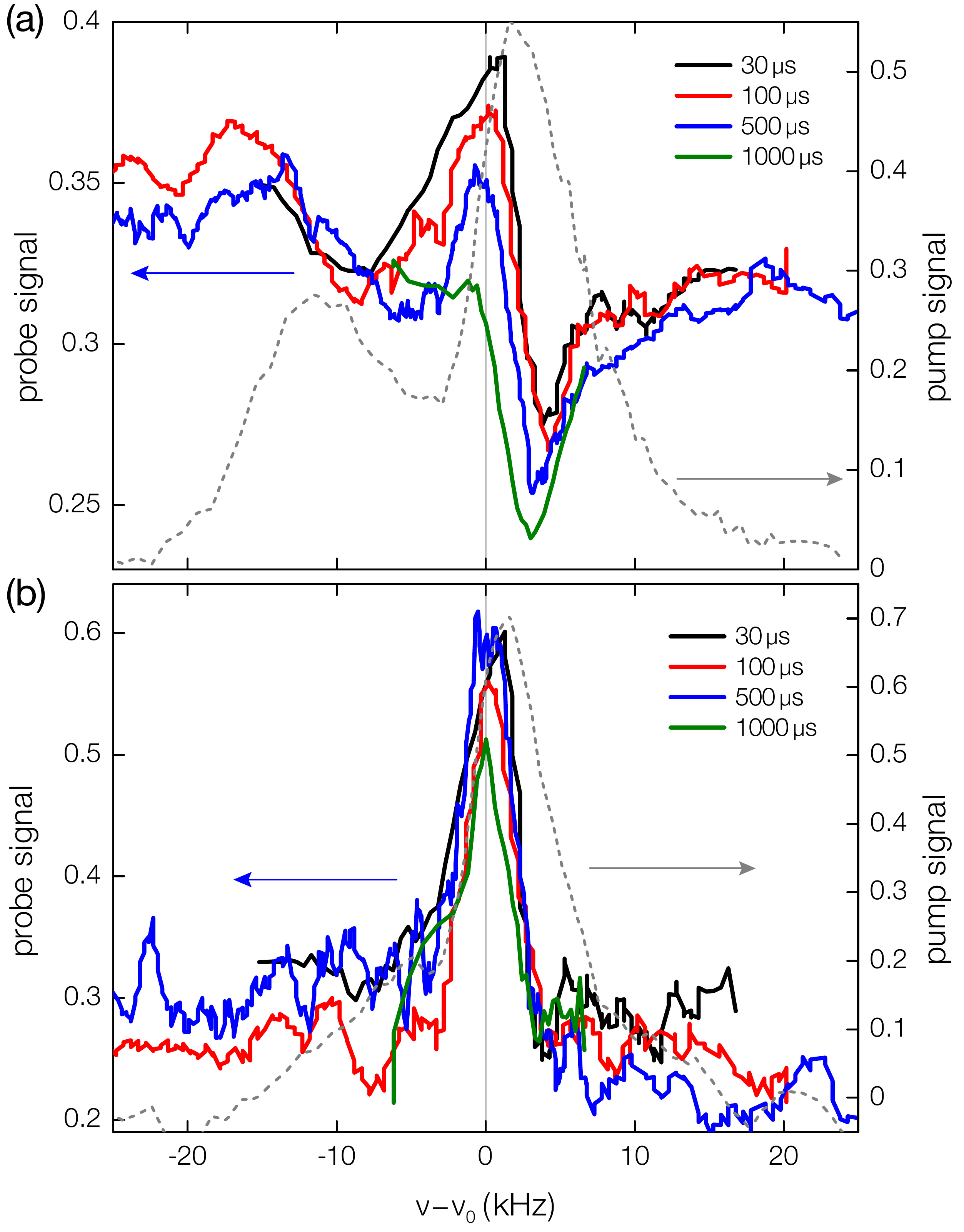}
\caption{
Spectral response of a  three-component 1-2-3 gas at various evolution times for $\kappa_F(0)\,a \sim$ 1.7 (see text), for the central region (a) and the outermost region of the trapped sample (b). In both panels, the pump spectra are shown as a reference (gray dashed lines, right $y$-axis), and the different colors correspond to the different evolution times indicated in the legend.
While state-3 atoms evolve into the emulsion phase within the central region, leading to $\Delta_+(t) \rightarrow 0$, state-2 atoms remain mixed at all times with all other particles (state-1 and 3 fermions, and 1-3 molecules, respectively). As such, the spectral response of state-2 atoms starkly differs from that of state-3 surviving fermions, simultaneously probed with the near resonant 2-3 RF frequency pulse. This leads, within the denser part of the sample, to a double-peak structure at positive detuning: the negative peak refers to state-2 atoms being transferred in  state 3. The positive peak near zero detuning corresponds instead to state-3 atoms transferred back from the emulsion phase.  The double feature is absent in the paramagnetic wings of the distribution where both state-3 and 2 atoms are always mixed with the surrounding medium. In panel (a) it is also visible the onset of the molecule dissociation spectrum at negative detuning, whose weight progressively increases with time.}
\label{2peaks}
\end{figure}
The outcome of this characterization, performed at different evolution times, is summarized in Fig.~\ref{2peaks}. In particular, Fig.~\ref{2peaks}a shows the signals $N_2 /(N_2 +N_3)$ recorded at various times (see legend) within the denser region of the sample, where the emulsion is expected to form, whereas Fig.~\ref{2peaks}b presents the spectral response obtained for the outermost, low density part of the cloud, likely remaining in the paramagnetic state.
For both cases, the spectroscopy signal $N_3 /(N_2 +N_3)$, measured by transferring the weakly-interacting 1-2 mixture into the strongly repulsive 1-3 one (dashed gray curve, right axis), is also shown as a reference. 

Focusing on Fig.~\ref{2peaks}a, associated with the central trap region, one can indeed notice the appearance of two oppositely-oriented spectral features: based on the previous discussion, we interpret the first positive bump close to zero-detuning as arising from state-3 atoms in a spin-polarized domain transferred into state 2 at the free atom resonance frequency. The second negative feature at positive detuning marks instead the optimum  frequency for transferring state-2 atoms, initially spread over the entire trap, into the repulsive regime. 
As the system evolves in time, the former peak becomes progressively lower owing to molecule formation, while the latter peak amplitude remains essentially unchanged. Parallel to the decreased weight of the $3 \rightarrow 2$ spectral contribution, one can also notice the emergence of the molecule dissociation spectrum at negative detunings.
Further, we remark that the opening of inelastic decay channels associated with the presence of three distinguishable fermionic species, although not completely impeding the development of the emulsion phase at short times, considerably decreases its lifetime relative to the case considered in the previous sections, where an almost balanced 1-3 atomic mixture was prepared.
In Fig.~\ref{2peaks}b we show how such a peculiar double-peak structure, revealed within the denser region of the sample, is totally
absent when monitoring the spectral response in the outer, low-density wings of the gas.
According with our expectation, owing to the reduced density, and hence interaction strength, this latter region of the sample remains in the paramagnetic phase at all times. This leads to the appearance of one single spectral feature at small positive detuning, weakly varying over 1\,ms of evolution.

In conclusion of this section, we remark that throughout our work we have discussed the properties of this highly-correlated state of fermions and pairs in terms of a heterogeneous quantum emulsion, consisting of micro-domains of size on the order of the interatomic distance. In some sense, this regime would not substantially differ from a paramagnetic atom-molecule mixture, where the wavefunctions of the three different components have vanishing overlap and are anti-correlated \cite{Zhai2009}.  Ultimately, to distinguish between those two scenarios requires either detailed many-body calculations, or experiments with microscopic probes at the atomic length scale.


\vspace{4mm}
\section{Conclusions and Outlook}
To summarize, in this work we have thoroughly explored the properties of a highly-correlated state emerging from a Fermi gas mixture coherently quenched to strong repulsion \cite{Amico2018}. 
Trap release measurements have enabled us to investigate how the correlated state trades in kinetic energy for interaction energy, and to reconcile seemingly discordant observations reported in previous studies \cite{Jo2009,Sanner2012}.
Rabi oscillation measurements unveiled in real time the breakdown of the homogeneous repulsive Fermi liquid, in favor of a rapidly developed heterogeneous phase composed of molecular and spin-polarized atomic micro-domains \cite{Amico2018}.
The regions of metastability for this quantum emulsion in terms of temperature and interaction strength have been tested through RF pump-probe spectroscopy. The micro-emulsion state is favored by low temperature and strong interactions, and consists of a metastable atom-molecule mixture which is not described by thermal and chemical equilibrium. 
Finally, while  micro-scale phase separation of atoms and pairs at strong coupling still lacks direct observation, strong signatures of the heterogeneous character of such a highly-correlated regime were obtained, both via the study of density-density correlations and through novel spectroscopic protocols.   
As such, our study provides new important insights in the compelling dynamics of a repulsive Fermi gas, subject to the concurrent action of pairing and ferromagnetic instabilities. While dissipative pairing processes do not fundamentally hinder the possibility to investigate some aspects of  Stoner's model within such an ultracold system, they appear crucially important for the emergence of a new form of quantum matter.

The heterogeneous character of the resulting metastable phase investigated in this work links the physics of the repulsive Fermi gas to certain strongly correlated electron materials, where competing order parameters coexist in nano-scale phase separation \cite{Dagotto2003,Dagotto2005}.
The realization of such an exotic atom-pair quantum emulsion opens unforeseen new perspectives: In the future, it will be interesting to explore the finite-momentum response and the transport properties of atoms and pairs in such a spatially inhomogeneous gas, and to explore its robustness in weak optical lattices \cite{Pilati2014, Zintchenko2016} or lower dimensions \cite{Conduit2010,Cui2014}. 
Further, quantum gas microscopes could uniquely explore the emergence of such a phase, with the competition between anti-ferromagnetic ordering favored by the undelying lattice structure and quantum emulsions of itinerant fermions.
Finally, our protocols could provide exciting possibilities to dynamically create elusive phases of magnetized superfluidity \cite{Casalbuoni2004, Fukushima2011, Bennemann2014}, and to spontaneously attain mesoscopic magnetic impurities within strongly interacting superfluids \cite{Magierski2019}.

\begin{acknowledgments}
We acknowledge insightful discussions with Stefano Giorgini, Dmitry Petrov and Alessio Recati. Special thanks to the LENS Quantum Gases group. 
This work was supported under European Research Council GA no. 637738 PoLiChroM, and no. 307032 QuFerm2D, and European Union Horizon 2020 research and innovation programme under the Marie Sk{\l}odowska-Curie GA no. 705269. 
\end{acknowledgments}

\end{document}